\documentclass[aps,showpacs,twocolumn,superscriptaddress]{revtex4}

\usepackage{graphicx}
\usepackage{dcolumn}

\def\slashchar#1{\setbox0=\hbox{$#1$}
   \dimen0=\wd0 \setbox1=\hbox{/} \dimen1=\wd1
   \ifdim\dimen0>\dimen1 \rlap{\hbox to \dimen0{\hfil/\hfil}} #1
   \else  \rlap{\hbox to \dimen1{\hfil$#1$\hfil}} / \fi}
\def\tstrut{\vrule height2.5ex depth0pt width0pt} % used in tables
\begin{document}
%\begin{flushright}
%GSI-Preprint-2003-15}
%\end{flushright}  
\title{Quark mass dependence of s-wave baryon resonances}
\author{C. Garc\'\i a-Recio}
\affiliation{Departamento de F\'\i sica 
Moderna, \\Universidad de Granada, E-18071
Granada, Spain}
\author{M.F.M. Lutz}
\affiliation{Gesellschaft f\"ur Schwerionenforschung (GSI),\\
Planck Str. 1, D-64291 Darmstadt, Germany}
\author{J. Nieves}
\affiliation{Departamento de F\'\i sica
Moderna, \\Universidad de Granada, E-18071
Granada, Spain}
%\today
\begin{abstract}
\rule{0ex}{3ex} We study the quark mass dependence of $J^P =
\frac12^-$ s-wave baryon resonances. Parameter free results are
obtained in terms of the
leading order chiral Lagrangian. In the 'heavy' SU(3) limit with
$m_\pi =m_K \simeq $ 500 MeV the resonances turn into bound states
forming two octets plus a singlet representations of the SU(3) group. A
contrasted result is obtained in the 'light' SU(3) limit with $m_\pi
=m_K \simeq $ 140 MeV for which no resonances exist. Using physical
quark masses our analysis suggests to assign to the $S=-2$ resonances
$\Xi(1690)$ and $\Xi(1620)$ the quantum numbers $J^P=1/2^-$.
\end{abstract}


\maketitle

%=========================================

%\tableofcontents

\section{Introduction}

The question what is the true nature of baryon resonances has
attracted considerable attention in recent modern constructions of
effective field theories describing meson-baryon scattering.  Before
the event of the quark-model it was already suggested by Wyld
\cite{Wyld} and also by Dalitz, Wong and Rajasekaran \cite{Dalitz}
that a $t$-channel vector meson exchange model for the s-wave
meson-baryon scattering problem has the potential to dynamically
generate s-wave baryon resonances upon solving a coupled channel
Schr\"odinger equation.  In a more modern language the $t$-channel
exchange was rediscovered in terms of the Weinberg-Tomozawa (WT)
interaction, the leading term of the chiral Lagrangian that reproduces
the first term of the vector meson exchange in an appropriate Taylor
expansion \cite{Wein-Tomo}. This offers a unique opportunity to study
the quark mass dependence of baryon resonances, one of the goals of
this work. Such studies may be useful to obtain a deeper understanding
of baryon resonances.  Here we follow a scheme proposed in
\cite{LK00,LK02}, based on the solution of the Bethe-Salpeter-Equation
(BSE), which incorporates two-body coupled channel unitarity, as other
approaches~\cite{weise}--\cite{grkl}, but also insists on an
approximate crossing symmetry.  Indeed the latter constraint led to a
parameter free description of the $\Lambda(1405)$ and N$(1535)$
resonances in terms of the WT Lagrangian~\cite{LK02}. The goal of this
paper is to systematically unravel the SU(3) structure of the lowest
lying s-wave baryon resonances. We show that two full octets plus an
additional singlet of resonances are dynamically generated within this
framework. In the SU(3) limit, with degenerate mesons and
baryons, such a structure was already found in Ref.~\cite{lambda}.

Of particular interest is the $S(Strangeness)=-2$ sector where we find
a narrow state (with a width of about 5 MeV) with a strong coupling to
$\bar K \Sigma $ suggesting the identification with the three star
resonance $\Xi(1690)$. For the latter resonance only its isospin
quantum number was established experimentally. Our analysis suggests
the quantum numbers $J^P=1/2^-$. This complements the conclusion of
the recent work of Ref.~\cite{Oset-prl}. The authors of this reference
also use a scheme based on the solution of the BSE with a kernel
determined by the WT term, and find just one resonance in the s-wave
$S=-2$ sector. The found resonance shows a large decay width and
Branching Ratios (BR) which are incompatible with the empirical
properties of the $\Xi(1690)$ resonance, and instead it was identified
with the one star resonance $\Xi (1620)$~\cite{Oset-prl}.  The main
difference between the approach of Ref.~\cite{Oset-prl} and that
followed here is the method used to renormalize the BSE. In
Ref.~\cite{Oset-prl}, a three-momentum ultraviolet cutoff of natural
size was introduced, though some channel dependence of its numerical
value was allowed.  Such a procedure turns out to work remarkably well
in the $S =-1$ sector at low energies, providing a good description of
the $\Lambda(1405)$ resonance, but it starts showing limitations at
higher energies, where the description of the $\Lambda(1670)$ and
$\Sigma(1620)$ is certainly poorer~\cite{benn}. Indeed, the procedure
of~\cite{Oset-prl,benn} does not work in the $S=0$ sector at all, and
it fails even to produce the lowest lying resonance
(N$(1535)$)~\cite{manolo}. Our scheme provides reasonable results in
the $S=0,-1$ sectors.  In the $S=-2$ sector,  we
also find, besides the resonance which we identify to the $\Xi(1690)$
and mentioned above, a resonance with the same features as that
described in \cite{Oset-prl}, which can be identified with the one
star $\Xi(1620)$ resonance.

%%   Fig. 1   %%%%%%%%%%%%%%%%%%%%%%%%%%%%%%%%%%%%%%%%%%%%%%%%%%%%%
%
\begin{figure}
\resizebox{0.4\textwidth}{!}{%
  \includegraphics{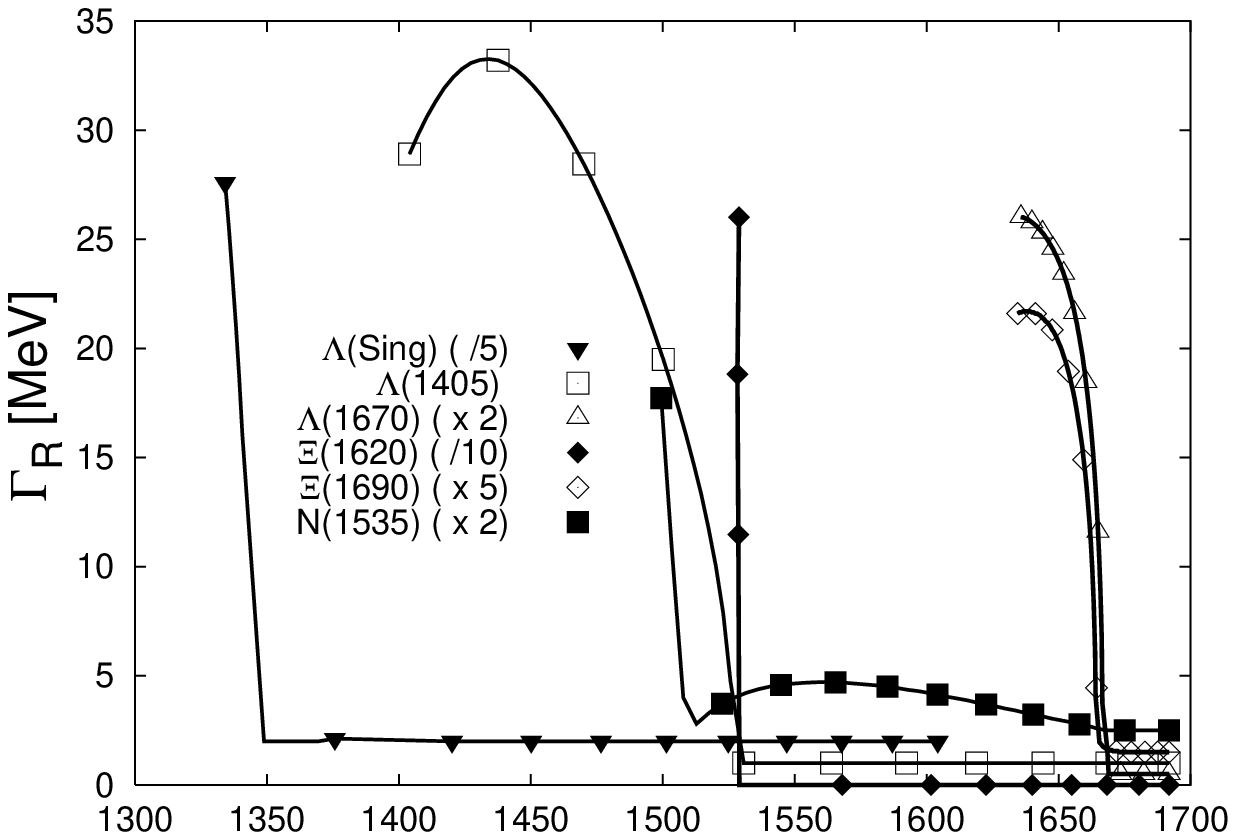}}
\resizebox{0.4\textwidth}{!}{%
  \includegraphics{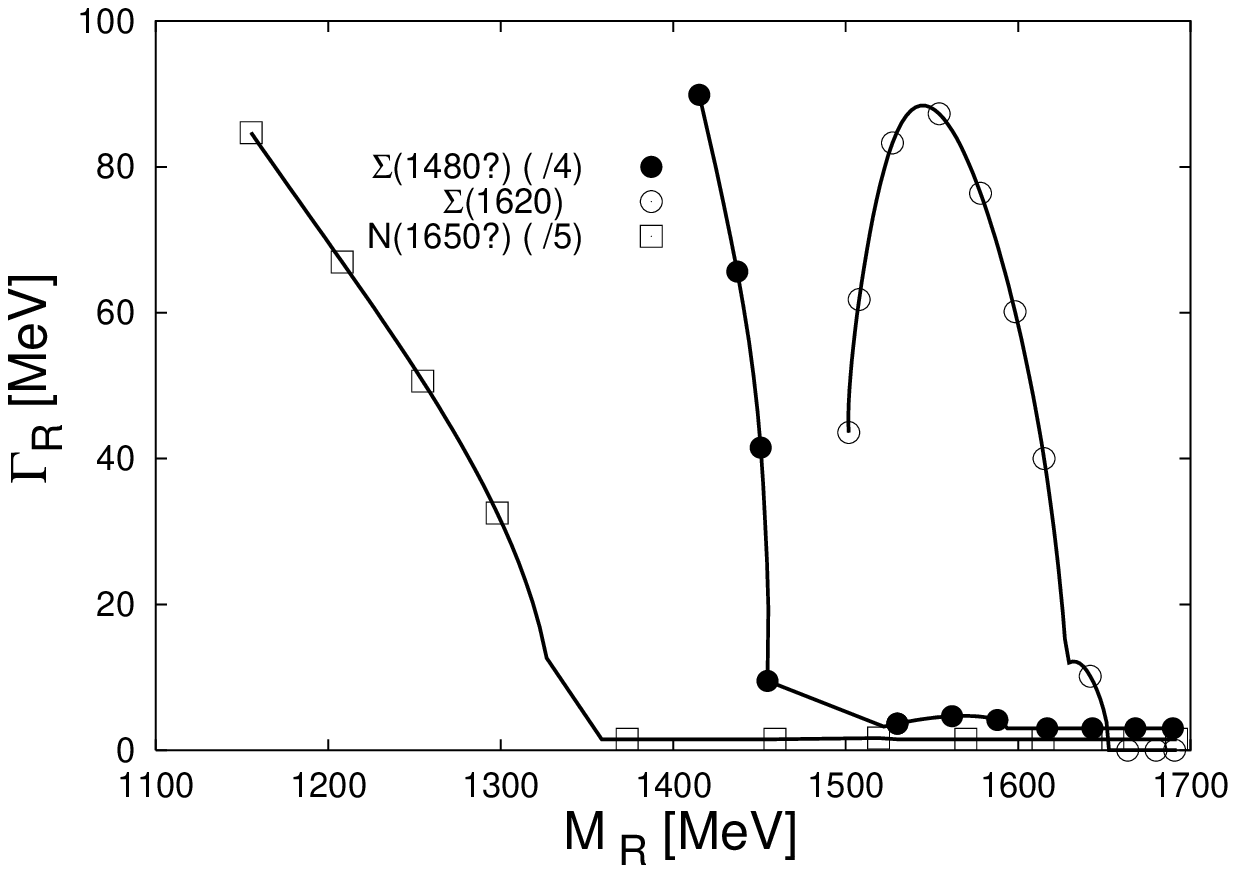}}
\caption{\footnotesize Masses and decay widths of two octets and a
singlet of baryon states for several values of the pion mass. For each
baryon state we plot eleven points, which correspond to eleven equally spaced
values of $x$ (Eq.~(\protect\ref{eq:su(3)})) ranging from 1 (first point
from the right) to 0 (first point from the left). Widths of different
baryon resonances have been scaled by factors, as it is  detailed in
the legend of the plots. To disentangle among different states, for
some of them, the widths have been shifted by constants
factors. Lines have been plotted just to guide the eye.} 
\label{fig:fig1}       % Give a unique label
\end{figure}

\section{Theoretical Framework}
\label{sec:th}
We solve the coupled channel BSE  with an
interaction kernel expanded in chiral perturbation theory as
formulated in~\cite{Pich95}. The solution for the coupled channel s-wave
scattering amplitude, $T(\sqrt{s})$ in the so called {\it on-shell}
scheme~\cite{LK02,EJmeson}, can be expressed in terms of
a renormalized matrix of loop functions, $J(\sqrt{s})$, and an
effective on-shell interaction kernel, $V(\sqrt{s})$, as follows
\begin{eqnarray}
T(\sqrt{s}) =  \frac{1}{1- V(\sqrt{s})\,J(\sqrt{s})}\,V(\sqrt{s})
\,. \label{eq:scat-eq}
\end{eqnarray}
Assuming the conservation of isospin and strangeness the scattering
problem decouples into 9 different sectors $(
(I,S)=(0,1),(1,1),(\frac12,0),(\frac32,0),(0,-1),(1,-1)$,
$(2,-1),(\frac12,-2),(\frac32,-2)$). In each sector, there are several
coupled channels, for instance, the $S=0$ sector requires four coupled
channels in the $I=1/2$ sector ($\pi$ N, $\eta$ N, $K \Lambda$ and $K
\Sigma$ ). The explicit form of the interaction kernel and the loop
functions can be found in \cite{LK02,grnpi,grkl}. The latter ones
logarithmically diverge and one subtraction is needed to make them
finite. Such a freedom can be used to incorporate approximate
crossing symmetry in the scheme, by the renormalization condition
\begin{eqnarray}
T(\sqrt{s}= \mu) = V(\mu ) \,, \qquad \mu = \mu (I,S) \label{eq:rsch}
\end{eqnarray}
where the natural choice
\begin{eqnarray}
&& \mu(I,+1)={\textstyle{1\over 2}}\,(m_\Lambda+ m_\Sigma) \,,
\quad \mu(I,0)=m_N\,, \quad  \label{eq:sub-choice}\\
&&  \mu(0,-1)=m_\Lambda,\,\, \mu(1,-1)=m_\Sigma, \,\,
\mu(I,-2)= m_\Xi \nonumber
\end{eqnarray}
is used as explained in detail in \cite{LK02}. It is evident that the
renormalization condition of Eq.~(\ref{eq:sub-choice}) is implemented
in a straight forward manner by imposing that the renormalized loop
functions $J(\sqrt{s})$ vanish at the appropriate points
$\sqrt{s}=\mu(I,S)$.  The renormalization condition reflects the fact
that at subthreshold energies the scattering amplitudes may be
evaluated in standard chiral perturbation theory with the typical
expansion parameter $m_K/(4 \,\pi f) < 1 $ with $f \simeq 90$
MeV. Once the available energy is sufficiently high to permit elastic
two-body scattering a further typical dimensionless parameter
$m_K^2/(8\,\pi f^2) \sim 1$ arises. Since this ratio is uniquely
linked to two-particle reducible diagrams it is sufficient to sum
those diagrams keeping the perturbative expansion of all irreducible
diagrams. This is achieved by Eq.~(\ref{eq:scat-eq}).  The subtraction
points of Eq.~(\ref{eq:sub-choice}) are the unique choices that
protect the s-channel baryon-octet masses manifestly in the p-wave
$J={\textstyle{1\over 2}}$ scattering amplitudes.  The merit of the
scheme \cite{LK00,LK02} lies in the property that for instance the
kaon-nucleon and antikaon-nucleon scattering amplitudes match at
$\sqrt{s} \sim m_\Lambda, m_\Sigma $ approximately as expected from
crossing symmetry. The subtraction points of Eq.~(\ref{eq:sub-choice})
can also be derived if one incorporates photon-baryon inelastic
channels in (\ref{eq:scat-eq}). Then additional crossing symmetry
constraints arise. For instance the reaction $\gamma \Lambda \to
\gamma \Lambda$, which is subject to a crossing symmetry constraint at
threshold, may go via the intermediate state $\bar K {\rm N}$.
Therefore the corresponding loop function must vanish identically at
$\sqrt{s}=m_\Lambda$ confirming Eq.~(\ref{eq:sub-choice}).  Here we
assume that this reaction is described by a coupled channel scattering
equation (\ref{eq:scat-eq}) where the effective on-shell interaction
kernel $V$ is expanded in chiral perturbation theory. We use the Leading Order
(LO) interaction kernel $V(\sqrt{s})$, as determined by the WT
interaction (see Refs.~\cite{grnpi,OR,Oset-prl}),
\begin{eqnarray}
V^{IS}_{ab}(\sqrt{s}) =
D^{IS}_{ab} \frac{2\,\sqrt{s}-M_a-M_b}{4\,f^2} \,,
\label{eq:lowest}
\end{eqnarray}
where $M_b$ ($M_b$) is the baryon mass of the initial (final) channel.
In Eq.~(\ref{eq:lowest}) tadpole terms, of subleading chiral order,
arising from the on-shell reduction of the interaction kernel (see
Ref~\cite{LK02,grnpi}) are neglected.  A parameter free prediction arises
if physical values for the meson and baryon masses are used. This is a
direct consequence of the chiral SU(3) symmetry of QCD that predicts
the strength of the WT interaction in terms of the parameter $f$ 
already determined by the pion decay process.

We will also study the quark mass dependence of the baryon resonances
that are dynamically generated by using meson and baryon masses that
deviate from their chiral SU(3) limit in Eq.~(\ref{eq:scat-eq}). We
use Goldstone boson masses as determined by the Gell-Mann,
Oakes and Renner (GOR) relation \cite{Pich95} in terms of the quark
condensate $\langle \bar u u \rangle = \langle \bar d d \rangle
=\langle \bar s s \rangle = -(280 \,\rm{MeV})^3 $, the current quark
masses $m_u=m_d = 3.5$ MeV and $m_s= 85$ MeV and $f=90$ MeV (with
these values we get $m_\pi=137.73$ MeV and $m_K=489.74$ MeV). The
masses of the baryon octet states are described in terms of the chiral
parameters $b_0=-0.346$ GeV$^{-1}$, $b_D = 0.061$ GeV$^{-1}$ and $b_F
=-0.195$ GeV$^{-1}$ (in the notation of Ref.~\cite{Pich95}, $b_{1,2}=
\mp (b_F \pm b_D)$). The above values require a baryon octet mass of
823 MeV in the chiral limit with $m_{u,d,s}=0$.  It is well known that
at this LO, all  baryons and Goldstone boson masses are
reproduced quite accurately (5\%).

\section{Results and Concluding Remarks}
%-------------------
\begin{table}[hb]
\vspace{-0.5cm}
\begin{center}
\begin{tabular}{c|c|c|c|c|c}
\hline
 $(I,S)$ & $M_R$  {[MeV]}  & $|g_i|^2$ & $\phi_i$
& BR$^{(\rm exp)}$& $g_i^{\rm b}$  \\\tstrut
Resonance [MeV] & $\Gamma_R$ {[MeV]} & & [Rad]
& [\%] &  \\\hline\tstrut
$(\frac12,0)$ &  & $[\pi $N]  0.1 & 1.1  & $45\pm 10$
&$-0.2$ \\\tstrut
 N(1535) **** &           & $[\eta $N] 4.7 &2.7     & $42\pm 13$
&$-1.6$ \\\tstrut
 $M=1505\pm 10$ & 1500     & $[K \Lambda]$ 4.2 & 6.2
& 0
&$ 0.7$ \\\tstrut
 $\Gamma=170\pm 80$ & 64     & $[K \Sigma]$ 11.4 & 6.0
& 0 &$ 2.5$ \\\hline\tstrut
$(0,-1)$ &   & $[\pi \Sigma]$  2.3 & 4.4  & 100 &$ 2.0$ \\\tstrut
 $\Lambda(1405)$ **** &          & $[{\bar K} $N] 9.3 & 0.3       & 0
&$ 2.0$ \\\tstrut
 $M= 1406\pm 4$ & 1409         & $[\eta \Lambda]$ 2.6    & 0.1       & 0
&$ 1.1$ \\\tstrut
 $\Gamma=50\pm 2$ &   34       & $[K \Xi]$ 0.1 & 4.3       &
0&$ 0.5$ \\\hline\tstrut
$(0,-1)$ &   & $[\pi \Sigma]$  0.04 & 1.9  & $40 \pm 15$
&$-1.3$ \\\tstrut
 $\Lambda(1670)$ **** &           & $[{\bar K} $N] 0.29 & 5.1       &
$25 \pm 5$
&$ 1.2$ \\\tstrut
 $M= 1670\pm 10$ &  1663          & $[\eta \Lambda]$ 0.99    & 3.4    &
$17 \pm 7$
&$-0.8$ \\\tstrut
 $\Gamma=35\pm 15$ &  12         & $[K \Xi]$ 9.69 & 0.1  & 0
&$ 2.4$ \\\hline\tstrut
$(0,-1)$ &   & $[\pi \Sigma]$ 8.2  & 5.7   & 100  &$ 2.4$ \\\tstrut
 $\Lambda(?)$ ? &           & $[{\bar K} $N] 5.0 & 2.2       & 0
&$-2.0$ \\\tstrut
 $M= ?$ &    1363    & $[\eta \Lambda]$ 0.5    & 1.6       & 0
&$ -1.4$ \\\tstrut
 $\Gamma=?$ & 115         & $[K \Xi]$ 0.3 & 5.5       &
0&$ 2.0$ \\\hline\tstrut
$(1,-1)$ &   & $[\pi \Lambda]$ 4.6 & 6.1  & seen &$ 0.8$ \\\tstrut
 $\Sigma(1620)$ ** &           & $[\pi \Sigma]$  3.1 &  0.6   &
seen  &$ 2.1$ \\\tstrut
 $M \approx 1620$ &  1505         & $[{\bar K} $N] 12.3 & 3.7
& $ 22 \pm 2$ &$-2.0$ \\\tstrut
 $\Gamma=87\pm 19$ &  21         & $[\eta \Sigma]$ 3.9  & 6.1    &
0&$ 0.8$ \\\tstrut
            & see [22]         & $[K \Xi]$ 0.5 &  3.5   &
0&$ 0.1$ \\\hline\tstrut
$(\frac12,-2)$ &   & $[\pi \Xi]$  7.5 & 5.6 & seen
&$ 2.6$ \\\tstrut
 $\Xi(1620)$ * &           & $[{\bar K} \Lambda]$ 5.2 &  2.8  & seen
&$-1.5$ \\\tstrut
 $M \approx 1620$ & 1565          & $[{\bar K} \Sigma]$ 0.7  &  2.6   &
0
&$-0.8$ \\\tstrut
 $\Gamma=23$ &  247        & $[\eta \Xi]$ 0.3      &  4.9   & 0&$ 0.3$ \\\hline
$(\frac12,-2)$ &  & $[\pi \Xi]$  0.02 & 0.1 & seen
&$-0.1$ \\\tstrut
 $\Xi(1690)$ *** &           & $[{\bar K} \Lambda]$ 0.16 &  6.0  &
seen
&$ 0.9$ \\\tstrut
 $M= 1690\pm 10$ & 1663          & $[{\bar K} \Sigma]$ 5.15  &  3.1   &
seen
&$-2.5$ \\\tstrut
 $\Gamma=10\pm 6$ &  4        & $[\eta \Xi]$ 2.28        &  3.2 & 0
&$-1.7$ \\\hline
\end{tabular}
\end{center}
\vspace{-0.4cm}
\caption{\footnotesize Experimental data, from Ref.~\protect\cite{pdg}
(PDG), and theoretical results for several resonances. When possible
we always quote PDG estimates for masses, widths and BR. If the latter
ones do not exist, we quote results from the most recent paper among
all quoted in~\protect\cite{pdg}. The following ratios, also given in
PDG, $\Gamma({\bar K} \Sigma) / \Gamma ({\bar K} \Lambda) = 0.75 \pm
0.39$ and $\Gamma(\pi \Xi) / \Gamma ({\bar K} \Sigma) < 0.09 $ for the
$\Xi (1690)$ are of interest, as well. For the case of the $N(1535)$
resonance, we quote the PDG pole position. From the theoretical side, we
look for poles in the SRS, as defined in \protect\cite{grnpi} (see
also note~[22]). Residues at the pole for each
meson-baryon channel, give the respective couplings and branching
ratios as defined in section II.D of \protect\cite{grkl} (note that the
$T$ matrix  define here (Eq.~(\protect\ref{eq:scat-eq})), coincides
with the $t$ matrix  used in \protect\cite{grkl}). Finally, in the last
column ($g_i^{\rm b}$) we give the couplings of the resonances to each
channel close to the heavy SU(3) limit ($x=0.98$, in Eq.~(\ref{eq:su(3)})),
where all of them become bound states. Besides to the states presented
in the table, in the heavy SU(3) limit there are two more bound
states: additional $N$ and $\Sigma$ states. Their couplings to the
different channels (we keep the ordering established in the table) are
(2.6,$-$0.5,1.5,$-$0.5) and ($-$1.1,1.5,0.7,$-$1.1,2.1) respectively.}
\label{tab:res}
\end{table}
We look for poles in the Second Riemann Sheet (SRS) of our amplitudes.
The positions of the poles determine masses and widths of the resonances,
while the residues for the different channels define the
BR~\cite{grnpi,grkl}. We find the resonances listed in
Table~\ref{tab:res} where only resonances with widths smaller than 250
MeV are included. Since for a resonance placed slightly above a
threshold the BR depends strongly on the exact position of the pole,
we only quote coupling constants (residues) that are much less
sensitive to the pole position. We find a remarkable success
predicting rather well the bulk of the features of the four stars
N(1535), $\Lambda (1405)$ and $\Lambda (1670)$
resonances~\footnote{There exists a second nucleon resonance
$(M_R=1156, \Gamma_R=415)$ MeV. Its large [small] coupling to the $\pi
{\rm N}$ [$\eta {\rm N}$] channel ($|g|^2_{\pi {\rm N}, \eta {\rm N},{
K} \Lambda, { K} \Sigma}=$ 8.5,0,0,3.3,0.3 ) and its SU(3) trajectory,
as we will see, make us to think that, despite its mass and width, it
might correspond to the four star N(1650) resonance. It is unclear
whether next-to-leading chiral corrections take the position of this
pole closer to that of the physical N(1650) resonance. A quantitative
description may require the inclusion of further inelastic channels,
like $\pi \Delta$ and $\rho N$ \cite{LWF02}.}.  We also find a
resonance in the $\Sigma$ channel, though its mass is a bit
small~\footnote{ We define a resonance as pole in an unphysical sheet,
usually the SRS (the SRS is determined by continuity to the First
Riemann Sheet [FRS]~\cite{grnpi}), with an appreciable influence into
the physical scattering line. In the $\Sigma$ channel, the pole
($M_R=$1505 MeV) listed in Table~\ref{tab:res} is above the first
three thresholds ($\pi\Lambda$, $\pi\Sigma$ and ${\bar K}{\rm N}$),
but it appears in the unphysical 11000 sheet, instead of in the 11100
one (each of the five digits counts for the number of turns around
each of the branch points~\cite{grnpi}). Between the third (${\bar
K}{\rm N}$) and fourth ($\eta \Sigma$) thresholds, the 11100 sheet
maps into the FRS, and thus is this sheet, the one which enters
into the definition of SRS of Ref.~\cite{grnpi}. Despite of this, it
is indeed the narrow ($\Gamma_R\approx 20$ MeV) pole located in the
unphysical 11000 sheet, with large couplings to the $\pi\Lambda$,
$\pi\Sigma$ and specially to the ${\bar K}{\rm N}$ channels, and
placed very close above the ${\bar K}{\rm N}$ threshold, the one which
has an important influence on the scattering for energies in the
neighborhood of the ${\bar K}{\rm N}$ threshold. Actually, the modulus
of the scattering matrix, for all open channels at these energies,
presents a peak, with an appreciable gap in the first derivative
(since the pole is placed above the third threshold (${\bar K}{\rm
N}$), but it is found in the 11000 sheet) which is clearly due to the
narrow pole listed in the table. There exists also a pole in the 11100
sheet ($M_R=1466, \Gamma_R=574$) MeV. It is above the ${\bar K}{\rm
N}$ threshold and has a large coupling to the $K\Xi$ channel. This
very broad pole is precisely the one quoted in Ref.~\cite{lambda}, but
it is placed so far from the scattering line, the $K\Xi$ threshold
($\approx 1810$ MeV) is also so far from the region of about 1500 MeV,
that it can not compete with the narrow one found in the 11000 sheet,
and it does not influence the physical scattering at all (chiral
corrections reduce its width and take its mass closer to the $K\Xi$
threshold indicating that it is the $\Sigma(1750)$
resonance~\cite{LK02}). There exists a third pole also in the 11000
sheet ($M_R=1446, \Gamma_R=343$) MeV, with a large coupling to 
$\pi \Sigma$, and located also just above the 
${\bar K}{\rm N}$ threshold. Its influence on physical scattering
processes is limited due to the presence of the narrower pole listed
in the table, but it might be identified to the $\Sigma (1480)$
bump~\cite{pdg}. }.  Besides, in the $S=-2$ sector we find two
resonances, which can clearly be identified to the $\Xi(1690)$ and
$\Xi(1620)$ resonances. Of particular interest is the signal for the
$\Xi(1690)$ resonance, where we find a quite small (large) coupling to
the $\pi\Xi$ (${\bar K}\Sigma$) channel, which explains the smallness
of the experimental ratio, $\Gamma(\pi \Xi) / \Gamma ({\bar K} \Sigma)
< 0.09 $~\cite{pdg} despite of the significant energy difference between the
thresholds for the $\pi \Xi$ and $ {\bar K} \Sigma$ channels. Thus,
this work widely improves the conclusions of Ref.~\cite{Oset-prl},
since we also address here the $\Xi(1690)$ resonance, and determine
its spin-parity quantum numbers ($J^P = \frac12^-$). On the other
hand, we also find a third $\Lambda$ resonance, not included in the
PDG yet, placed also below the ${\bar K} {\rm N}$ threshold and with a
large coupling to the $\pi \Sigma$ channel. This confirms the findings
of the recent work of Ref.~\cite{lambda}, where the chiral dynamics of
the two $\Lambda(1405)$ states is studied. The existence of the second
$\Lambda(1405)$ was firstly pointed out in Refs.~\cite{Ji01}
and~\cite{grkl}.

To explore the quark mass dependence of the resonances, we increase
the averaged $up$ and $down$ quark masses, but keep fixed the
antikaon mass. A parameter $x$ is introduced in
terms of which the pion mass varies as
\begin{equation}
\left. m_\pi^2\right|_{\rm SU(3)} = m_\pi^2 + x (m_K^2-m_\pi^2), 
\quad x\in[0,1] \,.
\label{eq:su(3)}
\end{equation}
A pair $(m_K^2,
\left. m_\pi^2\right|_{\rm SU(3)})$ determines the $\eta$ meson mass
via the GOR relation. Given the SU(3) symmetry breaking
parameters $b_0$, $b_D$ and $b_F$, the masses of the
baryon octet (N(940),$\Lambda(1115)$,$\Sigma(1190)$ and $\Xi(1320)$)
are also determined.  In the limit $x=1$ our SU(3) pion is as heavy as the
real kaon, while when $x=0$ the physical world is recovered, up to
some minor mass differences due to imprecisions of the GOR and baryon
splitting formulas. 

For the SU(3) symmetric $x=1$ case, where all baryons (mesons) have a
common mass $M$ ($m$), the $T$ matrix has poles in the FRS (bound
states). For each $IS$ channel, the position of the poles, $s_b$, is
such that the dimensionless function $\beta(s) =
2 f^2/\left(J(\sqrt{s})(\sqrt{s}-M)\right)$, at $s=s_b$, becomes an
eigenvalue of the real and symmetric matrix $D^{IS}$. The eigenvalues
of the latter matrices are 2,0,$-3,-3$ for both $IS=(1/2,0)$ and $IS
=(1/2,-2)$, and 2,$-6,-3,-3$ and 2,0,0,$-3,-3$ for $IS=(0,-1)$ and $IS
=(1,-1)$, respectively. Since $\beta(s)$ is negative between $(M-m)^2$
and $(M+m)^2$, only negative eigenvalues can be matched. Thus, we end
up, with two degenerate octets of mass $M_8=1691.83$ MeV (eigenvalue
$-3$, which has a multiplicity of two in all $IS$ sectors) and a
singlet of mass $M_0=1604.21$ MeV (eigenvalue $-6$ in the $\Lambda$
channel). Thus, we confirm here the findings of Ref.~\cite{lambda} on
the nature of the third $\Lambda$ listed in the
Table~\ref{tab:res}. Slightly away of the SU(3) symmetric world
($x\approx$0.98), we can determine the couplings of the baryon states
to each baryon-meson channel (last column of Table~\ref{tab:res}). The
sum of the squared of the couplings is given by
$g_{8,0}^2= \lambda_{8,0}
/\left(2M_{8,0}J(M_{8,0})\beta^\prime(M_{8,0}^2)\right)$, where
$\lambda_{8,0}=-3$ and $-6$ respectively. We find $g_8^2=9.65$ and
$g_0^2=15.83$. In Fig.~\ref{fig:fig1} we show the chiral behavior of
each of the members of the two octets and a singlet states. As
mentioned above, for $x=0$ one recovers the physical world, though the
results presented in the figure (specially for the widths) do not coincide
with those given in the Table~\ref{tab:res}. The differences are
produced by relative changes between the position of thresholds and
the location of the resonances, due to small differences
among the real meson and baryon masses (used in Table~\ref{tab:res})
and those predicted by the GOR and baryon splitting formula. Besides,
the identification in Fig.~\ref{fig:fig1} of the N$(1650)$  and the
$\Sigma(1480)$ bump is subject to all uncertainties discussed so far.
Finally in the 'light' SU(3) limit with $m_\pi
=m_K \simeq $ 140 MeV, the function $\beta(s)$, defined
above, is smaller than $-6$ in the whole interval $[(M-m)^2, (M+m)^2]$
and therefore bound states do no exist.

\begin{acknowledgments}
This
research was supported by DGI and FEDER funds, under contract
BFM2002-03218 and by the Junta de Andaluc\'\i a.

\end{acknowledgments}

%%%%%%%%%%%%%%%%%%%%%%%%%%%%%%%%%%%%%%%%%%%%%%%%%%%%%%%%%%%%%%%%%%%%%


\begin{thebibliography}{99}

\bibitem{Wyld} H.W. Wyld, Phys. Rev. {\bf 155} (1967) 1649.
\bibitem{Dalitz} R.H. Dalitz, T.C. Wong and G. Rajasekaran,
Phys. Rev. {\bf 153} (1967) 1617.
\bibitem{Wein-Tomo} S. Weinberg, Phys. Rev. Lett. {\bf 17} (1966) 616;
Y. Tomozawa, Nuov. Cim. {\bf A46} (1966) 707.
\bibitem{LK00}
M. F. M. Lutz and E. E. Kolomeitsev, Proc. of Int. Workshop XXVIII on Gross
Properties of Nuclei and Nuclear Excitations, Hirschegg,
Austria, January 2000.
\bibitem{LK02} M. F. M. Lutz and E. E. Kolomeitsev, Nucl. Phys. 
{\bf A700} (2002) 193; Found. Phys. {\bf 31} (2001) 1671.
\bibitem{weise} N. Kaiser, P.B. Siegel and W. Weise, Nucl. Phys. {\bf
A594} (1995) 325; Phys. Lett. {\bf B362} (1995) 23.
\bibitem{OR} E. Oset and A. Ramos, Nucl. Phys. {\bf A635} (1998) 99.
\bibitem{OM} J. Oller and U. Mei\ss ner, Phys. Lett. {\bf 
B500} (2001) 263.
\bibitem{grnpi} J. Nieves and E. Ruiz Arriola, Phys. Rev. {\bf D64}, 
(2001) 116008.
\bibitem{grkl} C. Garc\'\i a-Recio,
J. Nieves, E. Ruiz Arriola and  M. J. Vicente-Vacas, Phys. Rev.
{\bf D67} (2003) 076009.

\bibitem{lambda} D. Jido, et al., nucl-th/0303062.


\bibitem{Oset-prl} A. Ramos, E. Oset and C. Bennhold,
Phys. Rev. Lett. {\bf 89} (2002) 252001.

\bibitem{benn} E. Oset, A. Ramos and C. Bennhold,
Phys. Lett. {\bf B527} (2002) 99.

\bibitem{manolo} T. Inoue, E. Oset and M.J. Vicente-Vacas,
  Phys. Rev. {\bf C65} (2002) 035204.

\bibitem{Pich95} A. Pich, Rep. Prog. Phys. {\bf 58} (1995) 563.

\bibitem{EJmeson} J. Nieves and E. Ruiz Arriola,  Phys. Lett.
{\bf B455} 30 (1999);  Nucl. Phys. {\bf A679} 57 (2000).

\bibitem{pdg} K. Hagiwara et al., Phys. Rev. {\bf D66}, (2002) 010001.


\bibitem{Ji01} D. Jido, et al.,
Phys. Rev. {\bf C66} (2002) 055203.

\bibitem{LWF02}
M.F.M. Lutz, Gy. Wolf and B. Friman, Nucl. Phys. {\bf A706} (2002) 431.

\end{thebibliography}
\end{document}